# MEMS-based Speckle Spectrometer

A. I. Sheinis[a], L. Nigra[a] and M.Q. Kuhlen[b]

[a] Astronomy Department, University of Wisconsin, Madison, 475 N. Charter Street,
Madison
WI 53706

[b] Department of Astronomy and Astrophysics, UCSC, Santa Cruz CA 95064

**Abstract**

We describe a new concept for a MEMS-based active spatial filter for astronomical spectroscopy. The goal of this device is to allow the use of a diffraction-limited spectrometer on a seeing limited observation at improved throughput over a comparable seeing-limited spectrometer, thus reducing the size and cost of the spectrometer by a factor proportional to $r_0/D$ (For the case of a 10 meter telescope this size reduction will be approximately a factor of 25 to 50). We use a fiber-based integral field unit (IFU) that incorporates an active MEMS mirror array to feed an astronomical spectrograph. A fast camera is used in parallel to sense speckle images at a spatial resolution of $\lambda/D$ and at a temporal frequency greater than that of atmospheric fluctuations. The MEMS mirror-array is used as an active shutter to feed speckle images above a preset intensity threshold to the spectrometer, thereby increasing the signal-to-noise ratio (SNR) of the spectrogram. Preliminary calculations suggests an SNR improvement of a factor of about 1.4. Computer simulations have shown an SNR improvement of 1.1, but have not yet fully explored the parameter space.

Keywords, astronomy, speckle, spectroscopy, MEMS, adaptive optics, integral field spectroscopy

**Introduction**

Why is the space of the universe filled with galaxies?; Why are galaxies filled with stars?; Why are stars surrounded by planets?; Is the existence of life an extremely rare event or common in the universe?

We are beginning to answer some of these questions with existing telescopes on the ground and in space. The next generation of astronomical instruments will be based on extremely Large Telescopes (ELT's). These telescopes will be extraordinarily powerful tools for exploring the universe. They will see farther into space and farther back in time than any instruments currently in use and will give us unprecedented access to exquisite details of physical processes on both small and large scales and over most of the age of the universe. Several international groups are currently involved in the design of ELT's ranging in aperture from 25-50 meters.

As telescope apertures increase so do spectrometer apertures. In the era of the Extremely Large Telescopes this relationship drives seeing-limited high-resolution spectrometers to giant sizes and enormous costs. It is very desirable to develop techniques for spectroscopy that break the classical relationship between telescope and spectrometer size by allowing the use of a diffraction-limited spectrometer, whose size is independent of telescope aperture. To illustrate the size advantage, consider an R2 (63.4 degree grating) spectrometer for a 30-meter telescope with a reciprocal dispersion of 50,000 operating at a wavelength of 1 micron. The seeing-limited spectrometer requires a beam diameter of

1.84 meters (Bingham 1979), whereas the diffraction-limited spectrometer requires a 12.8 *millimeter* beam diameter.

This size differential illustrates the coming crisis in seeing-limited instrumentation for ELT's. It is unlikely that today's high-performance camera designs (i.e. LRIS (Oke et al 1994), ESI (Sheinis et al 1999), Binospec (Epps 1998) Imacs (Bigelow et al 1998) ) will be realizable in these apertures as the optical materials will not be available in large enough diameters. One solution is to continue to develop new and different optical designs to offset the growing aperture requirements, another is *to rethink the problem entirely and develop spectrometer designs whose size does not scale with the telescope size*. The latter solution is our long-term goal.

**System Description**

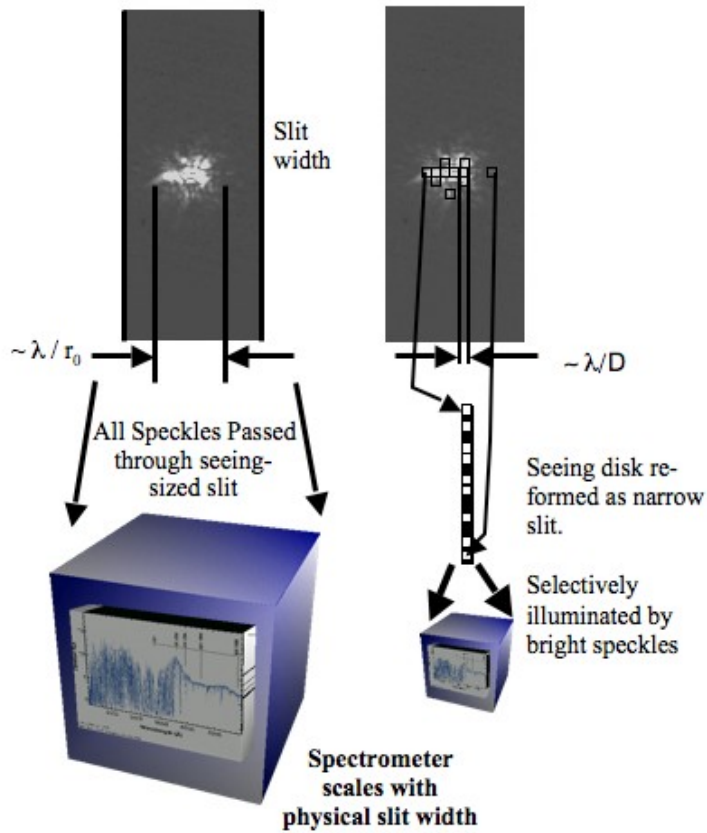

*Figure 1 illustrates the sensing concept. Rather than placing slit whose width is of order the seeing disk diameter $\lambda/r_0$ in the image plane as in a standard astronomical spectrograph, this new concept enables filtering of the image plane to send individual speckle images to the spectrograph. These speckle images have a width that is of order the diffraction limit of the telescope $\lambda/D$ (where D is the aperture diameter). Only speckle images that are above a preset threshold are redirected spatially and angularly to reside on a spectrograph entrance-slit of width $\lambda/D$. All light not in the speckles is redirected outside the spectrograph.*

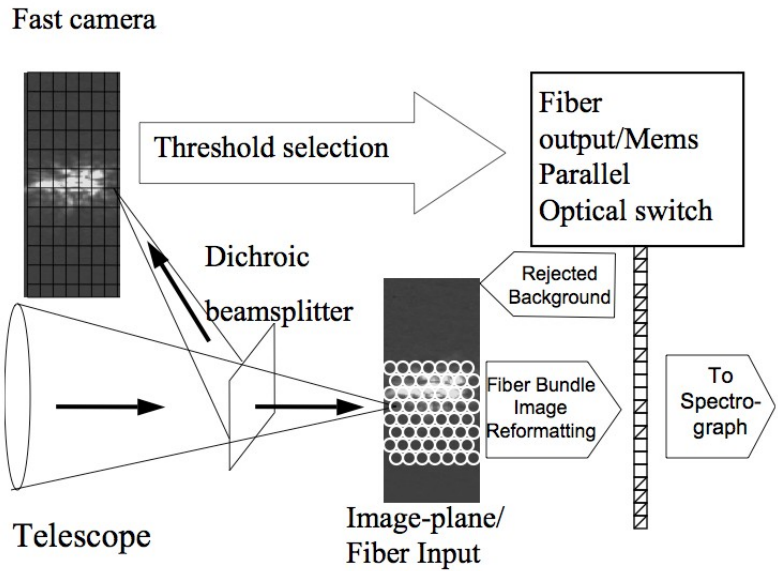

*Figure 2 - System concept. A Micro-electro-mechanical system (MEMS) (Vdovin et al 1995) parallel optical switch is placed at the entrance slit of an astronomical spectrograph. A dichroic filter is placed upstream of the image in order to divert a different band-pass to either image the speckle pattern on a fast CCD, or image the pupil onto a wavefront sensor. Telescope image scale is modified such that individual speckles are imaged onto (at least) one fiber input. Each fiber has an associated Mems switch channel. In this way, speckles above a preset SNR threshold are directed into the spectrograph. Speckles below the preset SNR threshold are dumped or reflected back into space. Integration times are long enough to reach a source-limited or sky-limited obnservations, with many cycles of the active spatial filter occuring within a single spectrometer integration. The spectrograph-camera magnification is such that the slit width $\lambda/D$ is (at least) critically sampled in the spectral direction by the spectrograph detector. In the spatial direction the spectrograph can be binned to reduce the read noise.*

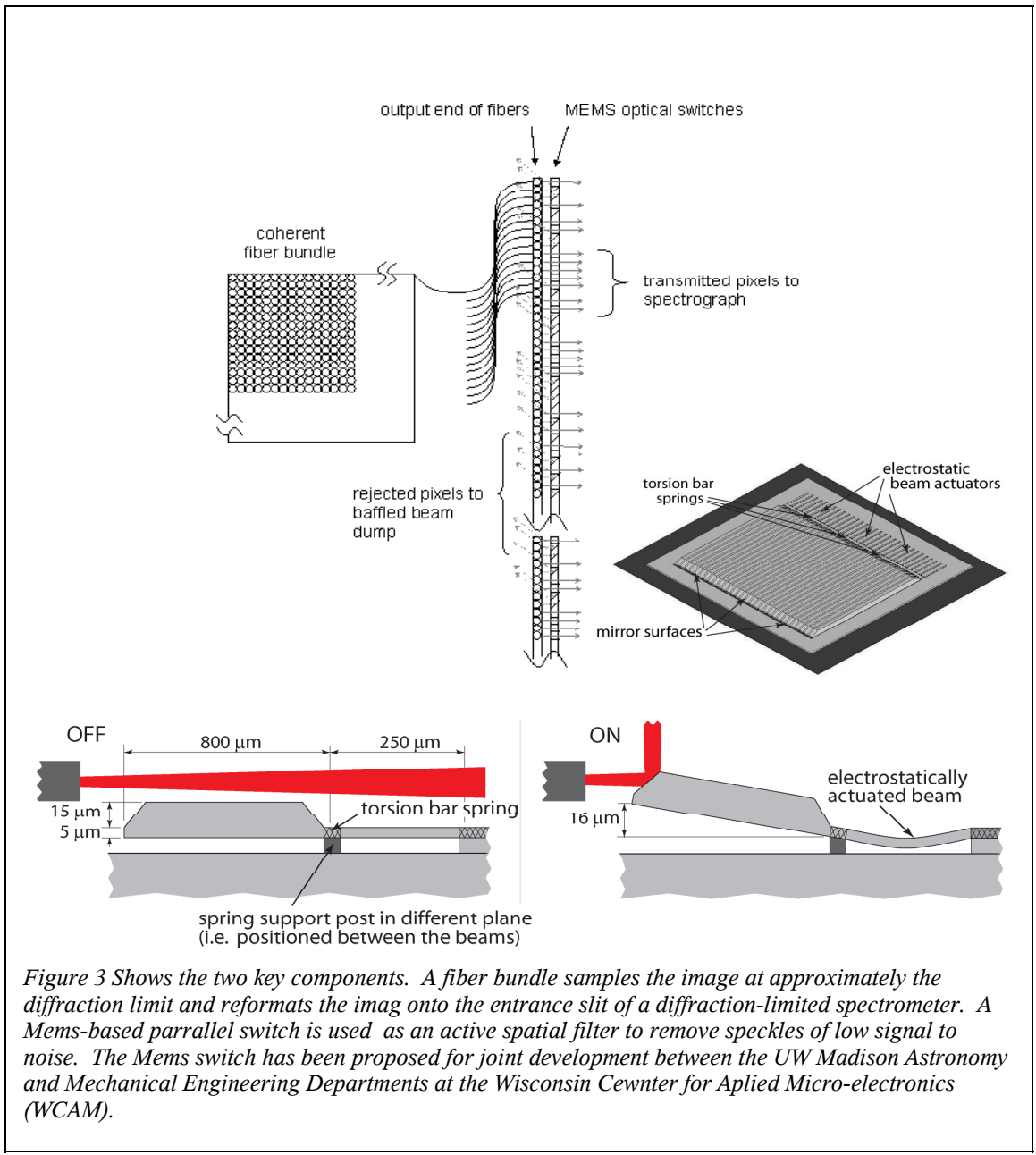

*Figure 3 Shows the two key components. A fiber bundle samples the image at approximately the diffraction limit and reformats the imag onto the entrance slit of a diffraction-limited spectrometer. A Mems-based parrallel switch is used as an active spatial filter to remove speckles of low signal to noise. The Mems switch has been proposed for joint development between the UW Madison Astronomy and Mechanical Engineering Departments at the Wisconsin Cewnter for Aplied Micro-electronics (WCAM).*

The development of near infrared adaptive optical (AO) systems for large ground based telescopes is actively being pursued by many groups. These systems are designed to both improve the spatial sampling and the signal-to-noise ration (SNR) in photometric and spectroscopic observations. Virtually all AO systems correct at or close to the pupil plane. The novel technique we propose is designed specifically to improve SNR of spectroscopic observations in the visible and near Infrared (IR) by correcting at the image plane. It does not, however, improve the spatial resolution of the image. It is therefore well-suited to objects whose angular size is less than the seeing disk diameter.

This new method uses a 2-D fiber bundle to reformat the image plane to a 1-D fiber array as shown in figures 2 and 3. The terminus of the fiber array is integrated with a MEMS parallel optical switch, such that the out put of each individual fiber can be directed either towards a slit spectrometer, or can be rejected by the switch to exit the system into a light baffle.

Two sensing options are being evaluated:

1) Thresholding: Since there is a one-to-one correspondence between the fast camera pixels and the pixels in the entrance slit to the spectrograph, a simple thresholding algorithm between the fast-camera signal and the MEMS mirrors will allow one to reject any pixels below a certain desired amplitude level from the spectrograph. This technique shows promise when sensing in a narrow band, close to the (narrow) spectroscopic band, or when a "guide star" is available.

2) Standard wavefront sensing: One can use a standard wavefront sensing technioque such as Shack-Hartman to sense at one wavelengtrh (i,e, vsisble) and reconstruct the speckle pattern at any other wavlength (i,e, NIR).

In this way, only image elements that contain a high signal-to-noise speckle will be passed to the spectrometer. The fast camera and MEMS switch operate at kHz, with mHz response time. The spectrometer detector integrates for periods on the order of hours.

## Current Issues

Technical issues to be addressed in this experiment:
1) Optical invariant or Etendue is set by the fiber array in order to avoid the scatter associated with using a bare spatial filter i.e. the bare TI DLP chip as operating on the image plane with a smaller than diffraction–limited spatial filter will scatter the beam into a large angle. Careful mating of the system magnification to the available fiber diameter will insure high throughput.
2) Scattered light off the closed fiber switch will require detailed modeling.
3) Fast camera integration time and MEMS switch control frequency will be optimized for high spectrometer SNR. This will depend on the details of the atmospheric turbulence.
4) Since the speckle pattern is a diffraction effect and thus function of wavelength, several sensing algorithms will be evaluated
    a. Dichroic sensing needs to optimize bandpass and proximity of sensing range to spectrometer range
    b. "Guidestar" sensing: A nearby bright star can be used in the same wavelength range as the spectrometer if it is within the 'isoplanatic" patch size of the atmosphere relative to the science star (Hardy 1998)
    c. "Standard" wavefront sensing: Using a Shack-Hartmann (Hardy 1998) or Curvature (Roddier 1999) wavefront sensor one can measure the wavefront directly and reconstruct the speckle pattern at any wavelength.
5) Another result of the speckle pattern scaling with wavelength is that as the bandpass of the optical filter increases the pattern is smeared radially, similar to lateral chromatic aberration. Optical designs for correcting optics that remove this radial wavelength dependence on the speckle image exist (Labeyrie 1995, Wynne 1979). We will evaluate these and compare to other solutions such as limiting the filter bandwidth.

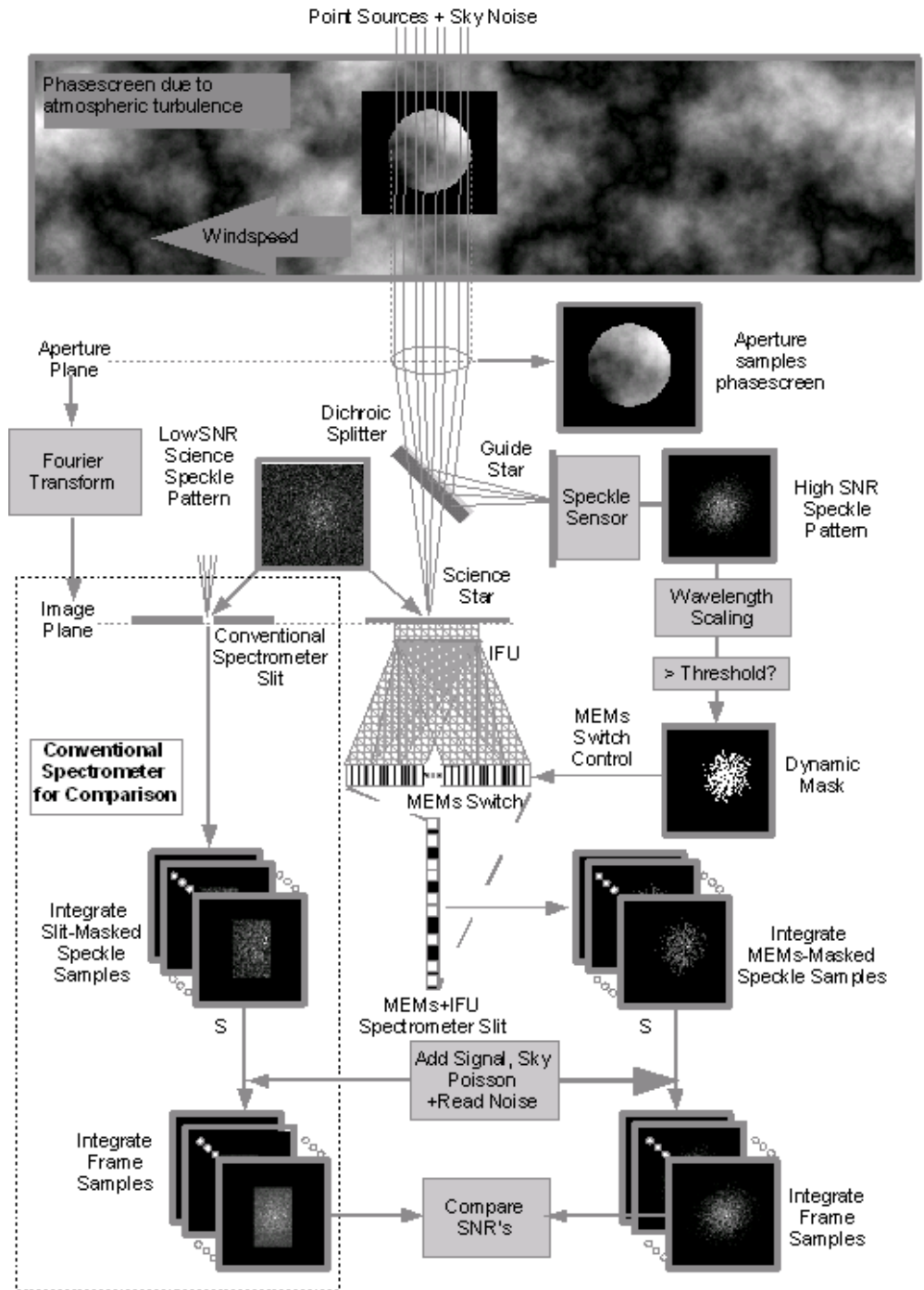

*Figure 4 - Schematic of the system and the STON.PRO IDL modeling package. Science and Guide starlight pass through a phasescreen produced from a Kolmogorov model of atmospheric turbulence which is sampled by the telescope aperture. Guide Star and Science Star separation is modeled as the equivalent of a dichroic splitter. Science star passes through a 2D-to-1D Integrated Field Unit (IFU) feeding the MEMs switch array controlled by Guide Star threshold logic on each sample to enable only high-SNR speckles. A number of MEMs-masked sample images are integrated to form a frame. Frames are read by adding Poisson and read noise and then summed through the simulation to obtain a Signal-to-Noise ratio. The Science Star is also passed through a conventional slitmask and otherwise processed identically in order serve as a reference for SNR performance comparison.*

## Modeling and Performance

In order to estimate the feasibility of this concept and to optimize it's design parameters, we produced a detailed computer-model of the signal and noise expected in such a speckle-slicing spectrograph. A set of IDL™ programs were written to model the MEMS-based spectrometer and compare its performance to a conventional slitmask spectrometer. The package processes co-field guide star and science star point sources through a Kolmogorov atmospheric turbulence model which simulates the behavior of the speckle pattern. The software is currently configured for point sources and the on-band guide star sensing option. Figure 4 is a functional diagram of the modeling software and is described below.

### Atmospheric Turbulence

The effect of the atmosphere is modeled as a time-varying Optical Transfer Function (OTF) resulting from a Kolmogorov turbulent layer passing across the field of view at a specified windspeed. The sky is uiformly illuminated by the distant source, so the OTF becomes the wavefront function. The model (Johansson 1976) produces a sample phase shift distribution function on the aperture plane whose width is that of the aperture and whose length depends on the windspeed and the exposure time. The phasescreen is a sample function reflecting the ensemble phase shift statistics of the Kolmogorov model. A sample phase screen function is (Johansson 1976).

$$s(x,y) = F^{-1}[N(k_x, k_y) \cdot \Phi(k_x, k_y)],$$

where

$F^{-1}[X(k_x, k_y)]$ is the inverse Fourier Transform,

$N(k_x, k_y)$ is a sample white noise spatial spectrum, and

$\Phi(k_x, k_y) = 0.023 \cdot r_0^{-5/3} \cdot k^{-11/3}$ is the phaseshift spatial spectrum .

Figure 4 includes a sample of the phasescreen, *s(x,y)*. The portion of the phasescreen sampled by the aperture at a given simulation time is interpreted as wavefront :

$$w(x,y) = p(x,y) \cdot e^{i \cdot s(x,y)},$$

where $p(x,y)$ is the pupil function at the aperture.

### Telescope

The telescope is simulated from aperture to image plane as the Fourier transform of the spatial distribution at the aperture to the angular distribution, which is in turn projected to the image plane based on the effective focal length of the instrument using the small angle approximation. This produces the speckle pattern.

Image plane sampling resolution is set by an oversampling parameter, which determines the number of pixels across the FWHM diffraction limit. From this, the angular and

spatial resolution on the image plane are established (see Table 1 for variable definitions):

$$\theta_{min} = 1.22 \cdot \lambda_0 / D \text{ the diffraction limit,}$$
$$\delta\theta = \theta_{min} / R_{os} \text{ the angular resolution per pixel (prior to downsampling),}$$
$$\delta p = \delta\theta \cdot F \text{ the physical image plane resolution per pixel.}$$

At the image plane, the guide and science star speckles are processed through separate optical paths.

**Guide star path**

Each guide star image sample is first spatially integrated (downsampled) to match the IFU resolution in the science star path. This image is then compared against a threshold to determine what portions of the science star image are to be passed through to the spectrometer during the sample period. This active masking is the essential feature of the device. The threshold can currently be determined by one of several simple algorithms. It can be proportional to the sky noise, a fixed value or a fraction of the guide signal. More sophisticated algorithms for determining this threshold to approach optimum performance are under investigation.

**Science star path**

The science star signal is processed in two ways; through a square slitmask followed by the MEMS device subsystem and, for comparison, through a rectangular slitmask only. After masking, both paths are integrated spatially in identical ways to simulate light collection resolution of the IFU. In the MEMS path, each IFU output sample is actively masked under control of the guide star speckle thresholding as described above, reducing the number of sky noise photons reaching the spectrometer during the exposure. In the comparison path, the IFU output is contunuously passed to the spectrometer throughout the exposure time.

Exposure time is split into a number of coherently integrated frames, each representing reading out a CCD at the same resolution as the IFU after a number of sample intervals. The frames are then summed to get the entire exposure.

**Noise**

Three noise components are included in the simulation; sky Poisson, signal Poisson, and readout. Throughout the simulation, the sky noise and signal photon distributions are identically processed, but separately maintained so that output SNR calculations can be performed after readout. At readout, the photon distributions are multiplied by a quantum efficiency factor and summed across all pixels to obtain total signal and sky electron counts. Sky and signal poisson noise components are computed from their respective counts as:

$$Noise = \sqrt{N_e}$$

Readout noise is modeled as a fixed number of electrons per pixel.

**Results**

In order to explore the parameter space, a controlling program sweeps the guide and science star magnitudes, the frame integration level and the IFU resolution. Figure 5 shows the results for a portion of this parameter space. Plotted are SNR improvement ratio (left axis) and actual SNR for a 1 hour observation (right axis) versus science star magnitude at a fixed guide star magnitude. Each curve is for varying speckle sensor pixel size, which also corresponds to IFU resolution. Table 1 summarizes the parameters used in generating the curves. Note that the simulated observation time of Figure 2 is actually 4.8 seconds and the SNR for 1 hour is calculated as the square root of the ratio.

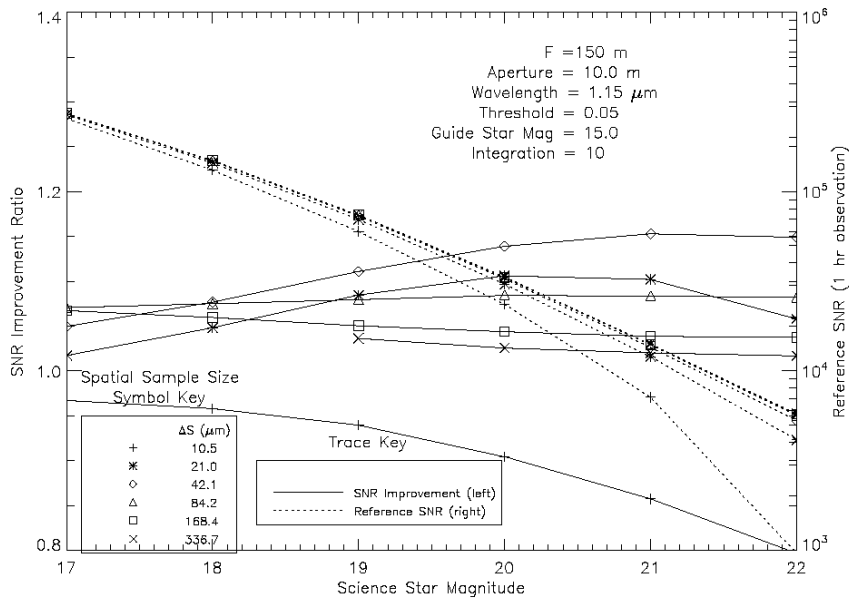

*Figure 5 - Simulated performance for parameters in Table 1.*

Note that when speckle sensing resolution is 10.5 microns (½ the diffraction limit) SNR degrades as the science star signal decreases (increasing magnitude). Here, the number of photons collected in the small pixel area is low and the threshold decision suffers from poor Poisson statistics allowing noisy pixels to be more likely enabled. As pixel size increases we begin to see SNR improvement which peaks at 42.1 micron, corresponding to twice the diffraction limit. This is approximately the nominal base width of a speckle. As the pixel size increases, the improvement lessens since the regions between speckles become less distiguishable. Here, low SNR regions are more likely to be enabled as they merge with high SNR regions in the speckle pattern.

For this parameter space, the best performance achieved is a SNR improvement ratio of 1.15, a 15% improvement in throughput. This occurs at science star magnitude 21 and resolution of 42.1 microns. Although this is significant, further improvement should be possible. It is expected that performance should approach SNR improvement of 1.4. The modeling program will be used to further explore the parameter space in order to get closer to this level of performance.

*Table 1 - Modelling Parameters*

| Description | Symbol | Value |
| --- | --- | --- |
| Aperture Diameter | D | 10 m |
| Focal Length | F | 150 m |
| Wavelength | $l_0$ | 1.15 mm |
| Fried's Parameter (turbulence scale) | $r_0$ | 45 cm |
| Seeing | $a_0$ | 0.643 arcsec |
| Diffraction Limit | $q_{min}$ | 28.9 milliarcsec |
| Conventional Slitmask Dimensions | N/A | 2 $a_0$ x 3 $a_0$ |
| Oversampling ratio | $R_{os}$ | 2 |
| Maximum image size (square) | $N_{pix}$ | 512 pixels |
| Image plane pixel (minimum angular) | dq | 14.5 milliarcsec |
| Image plane pixel (minimum physical) | dp | 10.5 mm |
| Aperture plane pixel | dx | 3.20 cm |
| Flux density reference ($m_J = 0$) | $F_{gl0}$ | 2.02 x $10^{10}$ photons/mm/sec/m$^2$ |
| Sky background intensity (J-band) | $I_{sky}$ | 19.6 Apparent mag / arcsec$^2$ |
| Sample time | $t_s$ | 40 ms |
| Coherent Integration | $N_{int}$ | 10 samples / frame |
| Non-coherent integration | $N_{blk}$ | 12 frames |
| Quantum efficiency | 0.75 | electrons / photon |
| Read noise | $n_r$ | 3 electrons / pixel |